\documentclass[traditabstract]{aa}
\usepackage{graphicx}
\usepackage{times}

\title{The solar differential rotation in the 18th century}
\titlerunning{Solar differential rotation in the 18th century}
\author{R. Arlt\inst{} \and H.-E. Fr\"ohlich\inst{}}
\institute{Leibniz-Institut f\"ur Astrophysik Potsdam (AIP), An der Sternwarte 16, D-14482 Potsdam, Germany}
\begin{document}

\date{Accepted \dots. Received \dots; in original form \dots}


\label{firstpage}

\abstract{
The sunspot drawings of Johann Staudacher of 1749--1799 were used to determine 
the solar differential rotation in that period. These drawings of the full disk 
lack any indication of their orientation. We used a Bayesian estimator to obtain 
the position angles of the drawings,  the corresponding heliographic 
spot positions, a time offset between the drawings and the differential 
rotation parameter $\delta\Omega$, assuming the equatorial rotation 
period is the same as today. The drawings are grouped in pairs, and 
the resulting marginal distributions for $\delta\Omega$ were multiplied. 
We obtain $\delta\Omega=-0.048 \pm 0.025$ d$^{-1}$ ($-2.75\,^\circ$/d) 
for the entire period. There is no significant difference to the value 
of the present Sun. We find an (insignificant) indication for a change 
of $\delta\Omega$ throughout the observing period from strong differential 
rotation, $\delta\Omega\approx -0.07~{\rm d}^{-1}$, to weaker differential 
rotation, $\delta\Omega\approx-0.04~{\rm d}^{-1}$.}


\maketitle
\section{Introduction}
The solar activity cycle is the result of an oscillatory 
magnetic dynamo in the interior of the Sun. Practically 
all available dynamo models for the Sun require a differential
rotation for part of the amplification of the magnetic fields.
Today's solar differential rotation is known from helioseismology
with fairly high precision down to the bottom of the convection
zone. 

The rotation period of the pole is considerably longer than 
the one at the equator. Balthasar et al. (1986) determined the
differential rotation from the rotational motion of sunspots.
Although sunspots provide no information on the rotational period at the pole,
we can extrapolate certain profiles of the rotation to the
pole and obtain a polar rotation period, which is 25\% longer than
the equatorial one. While the profile in Balthasar et al. (1986)
runs from about 24.7~d at the equator to 31~d at the pole, the
rotation profile obtained from helioseismology runs from about
25.2~d to about 38~d (Korzennik \& Eff-Darwich 2011). It is a
typical feature of solar activity that sunspot rotation periods
are slightly shorter than the rotation period of the photosphere (Benevolenskaya et al. 1999).


Several simulations of the dynamo process included the 
modulation of the differential rotation as a source of 
the solar cycle variability. Many global dynamo solutions are
kinematic solution, which do not integrate an equation of motion
over time. Small-scale motions are comprised  in a source term that usually
contains a nonlinearity that mimicks the back-reaction of the
magnetic fields on the flows ($\alpha$-effect and $\alpha$-quenching). Typically, the source term is
significantly diminished for magnetic fields $B$ stronger than
the equipartition field strength defined by $B_{\rm eq} = \sqrt{\mu_0\rho}\, u'$,
where $\mu_0$ and $\rho$ are the magnetic permeability and the
density, respectively, and $u'$ is the typical turbulent velocity. It is, however, not only the feedback to the small-scale motions but also the back-reaction on the large-scale flows, in particular the differential rotation, which introduces a non-linearity to the dynamo system. 

The fields generated in the dynamo process exert Lorentz 
forces, which modulate the differential rotation. This 
was studied by Malkus \& Proctor (1975) who found highly
nonlinear behaviour and complex time-series. These authors applied
the Lorentz force from the large-scale magnetic field to the
large-scale motion (differential rotation). In a refined attempt,
K\"uker et al. (1999) also included the back-reaction of the
generated large-scale field on the turbulent redistribution of 
angular momentum ($\Lambda$-effect and $\Lambda$-quenching). 
They found time-variations of the dynamo
solution that are very reminiscent of grand minima in the
solar activity time-series, e.g. the Maunder minimum. 

These models are of course accompanied by amplitude variations
of the differential rotation. It will therefore be extremely
interesting to measure the differential rotation at various
instances of the 400-year record of telescopic sunspot observations.

An attempt to determine the differential rotation during the
period of the Maunder minimum indeed indicated a differential
rotation that was stronger in the period of 1701--1719 than 
on the present Sun (Ribes \& Nesme-Ribes 1993). The whole
rotation period of the Sun appeared to be shorter, with a slow-down
at the turn of the 17th century.

We investigated the period of 1749--1799 with the attempt to 
determine the differential rotation of the Sun in a thorough 
statistical analysis.

\section{Data set}
We used the drawings of the full solar disk made by Johann Caspar 
Staudacher in Nuremberg in the second half of the 18th century as 
described by Arlt (2008). All drawings were made employing the 
projection method which can be deduced from the orientations of 
the drawings of solar eclipses. The drawings lack any indication 
of the orientation. The sunspot positions determined by Arlt (2009) were obtained by matching the solar rotation profile with the spots in adjacent drawings.  Here we cannot  take the sunspot positions obtained by Arlt (2009) into
account, since these employed the rotation profile by Balthasar
et al. (1986) to find the position angles of the solar disks, and
do not allow an independent determination of the differential rotation.

The entire set of sunspot drawings by Staudacher was examined in
suitable pairs, which contain several apparently identical spots 
in both images. It is obvious that we have to compile a considerable 
number $N$ of pairs in a sample to determine the differential
rotation. Apart from the intrinsic plotting errors in the drawings,
additional errors may occur when spots in adjacent drawings were
associated with each other, while they were actually not related.

\section{Data analysis}
\subsection{Defining the unknowns}

The entire 999 individual drawings of the solar disk were 
inspected manually for suitable pairs of drawings. Drawings 
with less than three spots were omitted, since they do not 
deliver  enough information to statistically determine 
of the free  parameters. A final set of 288 pairs were found 
with at least three spots in common. Several drawings were 
used twice if the combinations with the preceding and the 
succeeding  drawings provided suitable pairs. As in Arlt (2009), 
the drawings were cut out of the photos by defining the left, right, 
lower, and upper edge of the solar disk on the screen. Parts of 
any elliptical deformations can be compensated for in this way. 
The actual spot pairs were picked in an image that contained the 
two superimposed drawings, which were  now perfectly  circular  
and slightly magnified. The scale  in the disk centre  is 0.12$^\circ$ 
per pixel in heliographic coordinates. Using the rotation profile 
by Balthasar  et al. (1986), we estimated the expected displacements 
of spots. After one day a spot at 10$^\circ$ heliographic latitude 
becomes displaced by $0.63^\circ$ in longitude against a spot at 
30$^\circ$ latitude. This displacement corresponds to about 5~pixels 
and should be measurable. Obviously, it is difficult to detect the 
differential rotation signal in the observations because of these 
quite small displacements. It is the statistical multitude of drawings 
that delivers a measurement of the differential rotation with reasonable 
accuracy.

\begin{figure}
\centering
\includegraphics[width=0.48\textwidth]{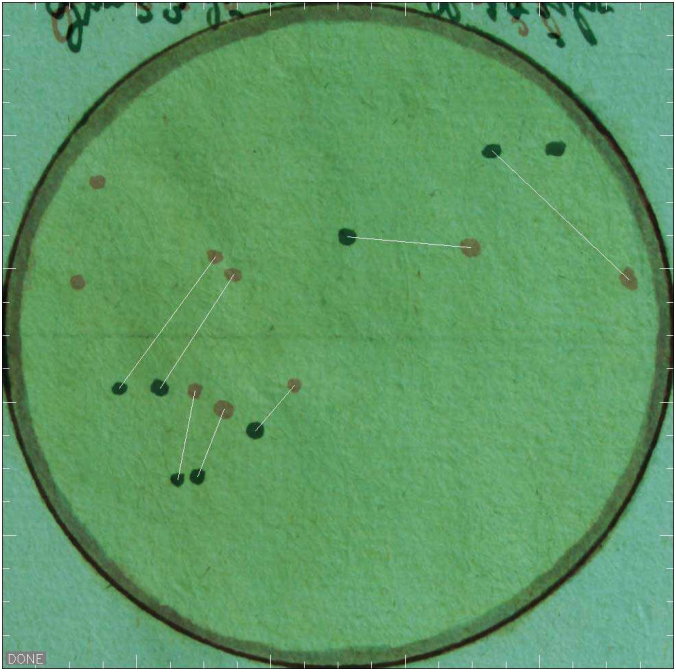}
\caption{Association of spots in the two superimposed images of 1761 June~23
(spots coloured in green) and 1761 June~24 (spots coloured in red).
The difference in position angles determined by rotational matching
is $29.8^\circ$.
\label{fdimage}}
\end{figure}

In any $i$-th pair of drawings, we measured the cartesian
coordinates $(x_{i,j}^{(1)}, y_{i,j}^{(1)})$ and $(x_{i,j}^{(2)}, 
y_{i,j}^{(2)})$ of $n_i$ spots visible in both drawing (1) and 
drawing (2), normalised in the solar radius, i.e. taking values 
between $-1$ and 1. We assumed that the longitudes $\lambda_{i,j}$, 
the latitudes $b_{i,j}$, and the two position angles of the drawings 
$p_i$, $q_i$ are unknown, where $j$ counts from one to the number of common spots $n_i$. 
We also assumed that the profile of the angular velocity can be expressed in 
terms of
\begin{equation}
  \Omega = \Omega_{\rm eq} + \delta\Omega \sin^2 b.
\end{equation}
In order to place a given spot at the same longitude in a pre-defined
rotating frame, we need to know the precise times of the two drawings.
Unfortunately, these are not always provided. Even when there are notes
about the times, they are hardly more precise than to the full hour. 
In principle, we can also keep the time difference $\Delta t_i$ between
the two drawings unknown. Note, however, that the three quantities
$\Omega_{{\rm eq},i}$, $\delta\Omega_i$, and $\Delta t_i$ are highly 
correlated. Any change in $\Delta t_i$ can always be compensated for by
a corresponding change in $\Omega_{{\rm eq},i}$.

In the same way, if the spots are on similar latitudes, the angular
velocity $\Omega_{{\rm eq},i}$ and the differential rotation parameter
$\delta\Omega_i$ are highly correlated, since a lower equatorial
velocity can be compensated for by a larger differential rotation
parameter. We have a few options to reduce this redundancy:
\begin{itemize}
\item We may assume that the times given by Staudacher are sufficiently
accurate and fixe $\Delta t_i$. The parameters $\Omega_{{\rm eq},i}$, 
$\delta\Omega_i$ have to reproduce the longitudinal motion of the spots
alone.
\item We may assume that the equatorial rotation period $\Omega_{{\rm eq}}$ of 
the Sun has not changed and fix it at today's value. If this cannot explain
the spots' longitudinal displacement, $\Delta t_i$ will compensate
for this. Remaining discrepancies can be mediated by $\delta\Omega_i$.
\item We may assume that the total angular momentum of the Sun has not
changed since Staudacher. If we assume that the shape of the internal
rotation profile has not changed (except its amplitude), and if we
assume that we know this profile precisely, we can find a relation
for $\delta\Omega_i(\Omega_{{\rm eq},i})$. Hence, only $\Omega_{{\rm eq},i}$
and $\Delta t_i$ are the unknowns. 
\end{itemize}

\begin{table}
\caption{Comparison of the number of measurements versus the number 
of free parameters for given spot numbers and different methods using 
pairs of drawings. The maximum number of common spots that appeared in 
Staudacher's drawings is 20.}
\label{unknowns}
\begin{tabular}{lcc}
Spots & Measurements & Parameters\\
\hline\\
2 & 8 & 8\\
3 & 12 & 10\\
4 & 16 & 12\\
5 & 20 & 14\\
6 & 24 & 16 \\
7 & 28 & 18 \\
\dots & \dots & \dots \\
20 & 80 & 44\\
\hline
\end{tabular}
\end{table}

The first option is the most natural choice but turned out to be 
problematic, since the times given by Staudacher appear to be rough 
estimates only. There is also a considerable number of drawings for 
which no time is given except the date. Excluding these would leave 
us with too few pairs to obtain a meaningful result. The last option 
appears to be the most physical one since it is highly unlikely that 
the total angular momentum has changed over 250~yr. The solar wind 
carries away only about $10^{30}$ g\,cm$^2$ s$^{-2}~{\rm sr}^{-1}$ 
(Pizzo et al. 1983). In 250~yr this is less than $10^{-8}\, A_\odot$, 
where $A_\odot$ is the total angular momentum of the Sun of about 
$A_\odot = 2\cdot 10^{48}$ g\,cm$^2$ s$^{-1}$ (e.g. Komm et al. 2003). 
The problem with this approach, however, is that fixing the total 
angular momentum does not fix the distribution of $\Omega$ at the 
surface well enough because of the low density and the less well 
known internal rotation profile of the Sun, which enters the total 
angular momentum. This is the reason why we chose a compromise of 
keeping the observing time open ($\Delta t_i$ is a free parameter) 
but fixed the rotation period at the equator to the present value. 
Because the specific angular momentum depends on the distance of a 
fluid element from the rotation axis, a change of $\Omega(b)$ at a 
given total angular momentum means a considerable change at high 
latitudes, but only a small change at the equator. Assuming a constant 
$\Omega_{\rm eq}$ is therefore a fair approximation of a constant total 
angular momentum.

From the two drawings of each pair and the two measured coordinates, 
we have $4n_i$ measurements for any given pair. The heliographic 
positions of these spots, the differential rotation parameter, and 
the time difference lead to $2n_i +4$ unknowns. Table~\ref{unknowns} 
gives an impression of the number of measurements versus the number 
of free parameters.

\subsection{Bayesian inference}\label{3.2}
We used a Bayesian approach to determine the differential rotation. 
The method is based on the studies by Fr\"ohlich et al.\ (2009) and 
Frasca et al.\ (2011), who determined the differential rotation of 
stars from photometric data. The idea is to define a model with free 
parameters and to determine the likelihood for many sets of parameters. 
The probability for each realisation of the model to have produced the 
data  is computed. There is yet another freedom in the system, which 
describes the uncertainty of the measurements. We have set this to a 
fixed value of 0.05~solar radii for the standard deviation, assuming 
Gaussian errors in all our runs. It is to represent the plotting 
accuracy when the spots are drawn using the projection method.

A main concern of Bayesian parameter estimations is to obtain a 
comprehensive idea of the probability distribution in the whole
parameter space. We therefore used a likelihood function for the 
$i$-th pair of drawings,
\begin{eqnarray}
\lefteqn{\Lambda(x_j, y_j; \lambda_j, b_j, \delta\Omega, \Delta t)=}\nonumber\\
&& \prod_{j=1}^{n_i} \frac{1}{\sqrt{2\pi} \, \sigma}
 {\rm exp} \left[- \left(\frac{\left[x_j - f_x(\lambda_j, b_j, \delta\Omega, \Delta t)\right]^2}{2\sigma^2} +\right.\right.\nonumber\\
&& \quad\quad\quad\quad\quad\quad \left.\left. + \frac{\left[y_j - f_y(\lambda_j, b_j, \delta \Omega, \Delta t)\right]^2}{2\sigma^2}\right)\right],
\label{product}
\end{eqnarray}
(omitting the indices $i$ for the sake of legibility) where $f_x$ 
and $f_y$ compute the theoretical cartesian coordinates of all $n_i$ 
common spots in the normalised solar disk from the set of parameters. 
In principle, we seek the probability distribution of a $\delta\Omega$ 
common to {\em all} pairs of drawings in the investigated period. This 
problem would create a huge parameter space with the entire heliographic 
spot coordinates, all position angles and all time offsets plus a single 
parameter $\delta\Omega$. Three spots in three images would then deliver 
18~measurements versus 12~unknowns.

Such a problem is computationally extremely demanding. Fortunately, the 
problem splits into subsets of parameters. These subsets  would be 
completely independent if the spots in one pair are different from the 
spots in the following pair. Then the probability distributions of the 
two individual parameters scans for the two  pairs can simply be multiplied. 
In practice a number of spots were ``reused'' in a following pair and thus 
lead to dependence  on the previous pair. In principle, we could have 
used longer ``chains'' of observations. However, because sunspot groups 
evolve, it becomes arbitrary to associate common spots over more than two 
drawings. Problems may be introduced by erroneously marking spots as being identical 
in different drawings that are in fact not. We therefore decided to restrict 
the analysis to pairs of drawings and to use the product of all 
probability distributions obtained from them for the final estimate of 
$\delta\Omega$.

\begin{figure}
\centering
\includegraphics[width=0.485\textwidth]{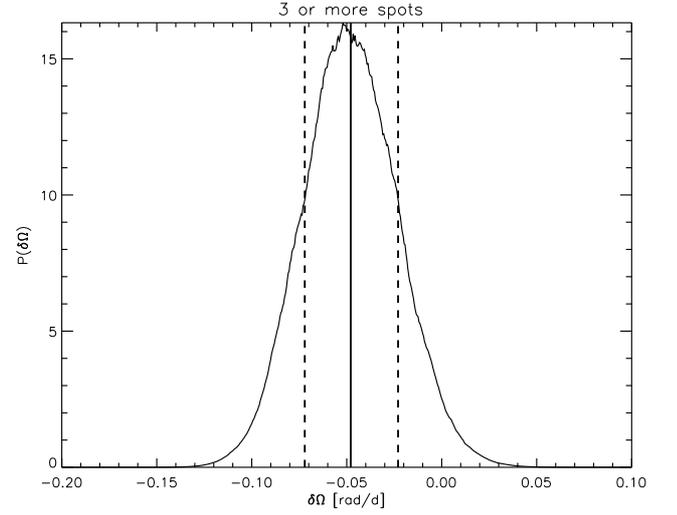}
\caption{Marginal distribution for the differential rotation
parameter for 
a minimum of three spots. A total of 156~pairs were used. The 
average value is $\delta\Omega = -0.048^{+0.025}_{-0.024}~{\rm d}^{-1}$.
\label{pca_spots3}}
\end{figure}

\begin{figure}
\centering
\includegraphics[width=0.485\textwidth]{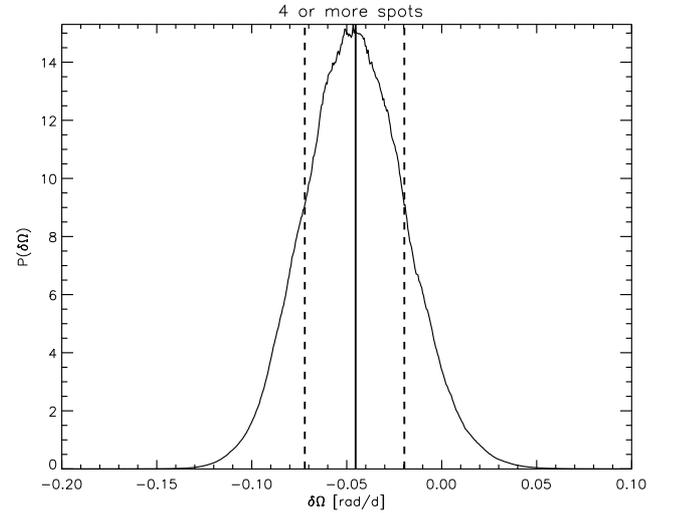}
\caption{Marginal distribution for the differential rotation
parameter for a 
minimum of four spots.  A total of 119~pairs were used. The 
average value is $\delta\Omega = -0.045^{+0.026}_{-0.027}~{\rm d}^{-1}$.
\label{pca_spots4}}
\end{figure}

All possible heliographic longitudes from $-\pi$ to $\pi$ were taken 
into account as well as all possible latitudes, but in the $\sin b$ 
space to account for the smaller areas covered by given latitude bands 
at high latitudes. The range of $\delta\Omega$ runs from $-0.5$ d$^{-1}$ 
to 0.5 d$^{-1}$ and includes an increase of angular velocity with 
latitude as well as even a pole rotating in the opposite direction from 
the equator. While the latter situation is highly unlikely, we did not 
want to restrict the parameter space to certain solutions. This is not 
a waste of computing power since the Markov chains will practically 
never enter this region of the parameter space because of the extremely 
small likelihood. The time offset is bound to $-0.6~{\rm d} \leq 
\Delta t_i \leq 0.6~{\rm d}$. The lower limit represents the situation 
of Staudacher observing near sunset on the first day and near sunrise 
on the second day. Formally, this corresponds to observing times at 
19$^{\rm h}$12$^{\rm m}$ on the first day  and at 4$^{\rm h}$48$^{\rm m}$ 
on the second day, assuming symmetry around midnight. The upper limit for 
$\Delta t_i$ is the opposite extreme case with an observation at 
4$^{\rm h}$48$^{\rm m}$ on the first day and at 19$^{\rm h}$12$^{\rm m}$ 
on the second day. The extreme sunrises and sunsets at Nuremberg are at 
3$^{\rm h}$52$^{\rm m}$ and 20$^{\rm h}$10$^{\rm m}$, respectively. These 
are not fully covered by our $\Delta t_i$ range but it would only be relevant 
if {\em both} extremes were used for two consecutive observations and only 
during a few weeks during the summer.

The full scan of a parameter space with more
than a few dimensions is technically not possible. An efficient tool 
for obtaining a representative idea of the probability distribution 
are Monte-Carlo Markov chains (MCMC, cf.\ Press et al.\ 2007). The 
enormous advantage of such a
search is that one obtains confidence intervals for all parameters
within the assumption of the model. Typically 64~Markov chains are
``exploring'' the parameter space in 20~million steps. 
In order to enhance the efficiency, MCMC was performed in
an orthogonalized space, in which the new variables, a linear
combination of the original parameters, were decorrelated.
A principal component analysis (PCA) was performed using
singular value decomposition (SVD, cf. Press et al. 2007) at
several instances during the Markov chain walks. While the
classical PCA relies on the covariance matrix and provides as
eigenvalues the square of the standard deviations, the SVD
results in standard deviations themselves. An SVD is to be preferred
because it is numerically more robust and faster (see e.g.
Wall et al. 2003 for more details). The background mathematics
behind the SVD method used is very complex and beyond the scope
of this paper. Note that a full decorrelation can only be obtained
if the probability distribution is a multivariate Gaussian; for
all other distributions, SVD (or PCA) does reduce correlations
between parameters, but does not make them vanish completely.

\begin{figure}
\centering
\includegraphics[width=0.485\textwidth]{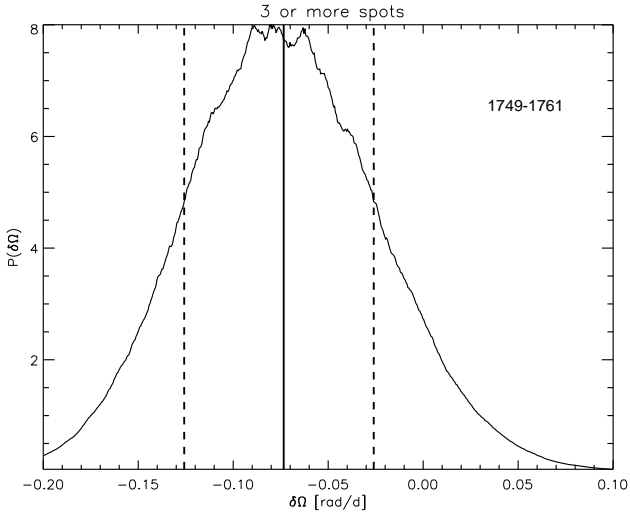}
\caption{Marginal distribution for the differential rotation
parameter for a
minimum of three spots. A total of 55~pairs were used. The average value is 
$\delta\Omega = -0.073^{+0.047}_{-0.052}~{\rm d}^{-1}$.
\label{pca_spots3_1749-1761}}
\end{figure}

\begin{figure}
\centering
\includegraphics[width=0.485\textwidth]{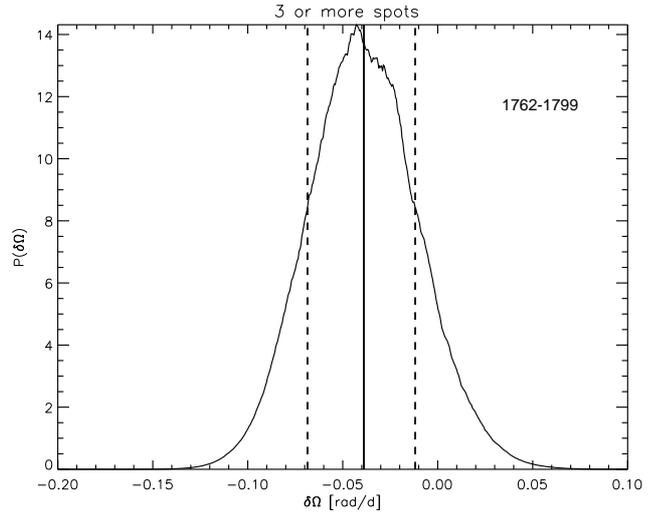}
\caption{Marginal distribution for the differential rotation
parameter for 
a minimum of three spots. A total of 101~pairs were used. The 
average value is $\delta\Omega = -0.039^{+0.027}_{-0.030}~{\rm d}^{-1}$.
\label{pca_spots3_1762-1799}}
\end{figure}

The set of parameters finally adopted was derived from the marginal
distribution, which is the probability distribution of one parameter
when integrated over all other parameters. Because we obtained the
probability distributions from $N$ pairs of drawings, we multiplied
all distributions to obtain a total distribution for the 
differential rotation parameter of the Sun. The result tells 
us for a given model the probability of each configuration
of this model to have produced the data.

\section{Results and discussion}
The probability distributions of all the 288~pairs were computed based 
on almost $1.3\cdot 10^9$ Markov chain steps for each of the pairs. 
Several distributions show peculiarities, such as a maximum at one of 
the edges of the considered parameter range. The reason for this is 
the association of spots in two drawings that have actually nothing in 
common. To exclude these unsuitable distributions, we selected only 
those for which the maximum probability lies inside the the tested 
intervals for $\delta\Omega$ and $\Delta t$ and not on the boundaries 
set in Sect.~\ref{3.2}. We also excluded totally flat probability 
distributions by setting a maximum standard deviation of a (supposed) 
Gaussian as a selection criterion.

\begin{figure*}
\centering
\includegraphics[angle=90,width=1.0\textwidth]{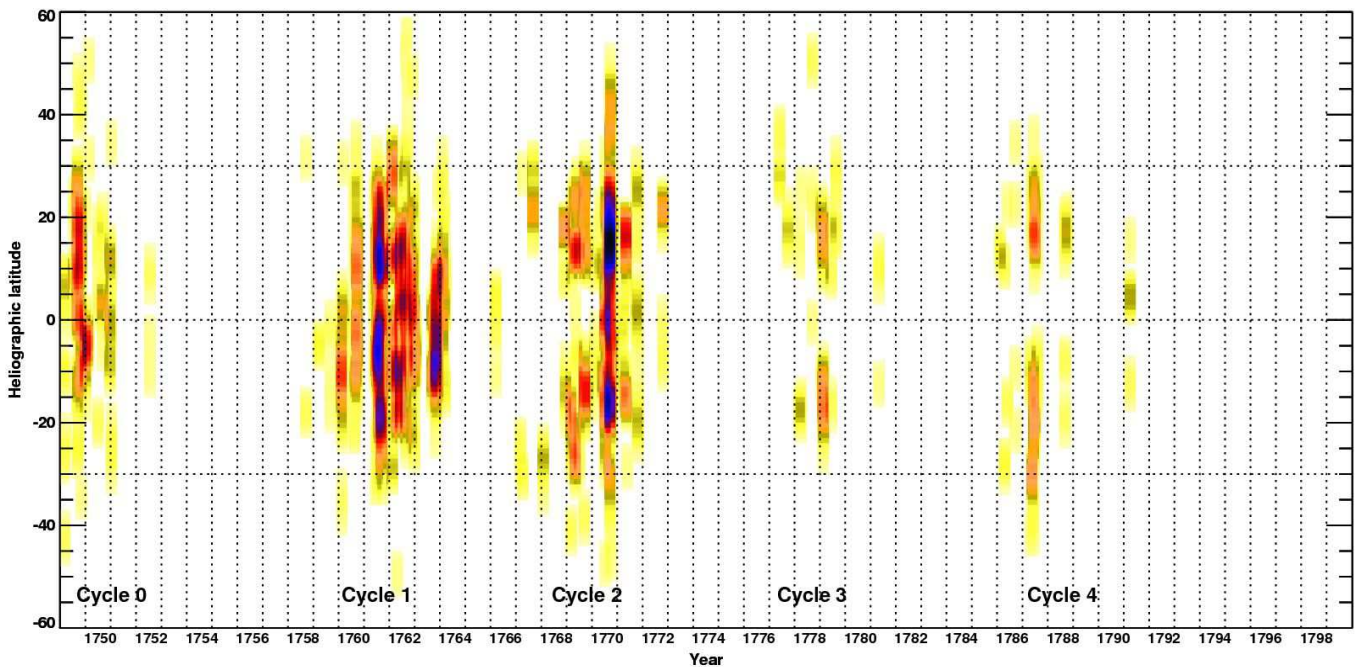}
\caption{Butterfly diagram from 1512~spot latitudes determined through the process
of Bayesian inference of the differential rotation. Individual spots were broadened
as if they had persisted for 80~days to obtain a smoother picture.
\label{b_pca_20m_jpg}}
\end{figure*}

Figure~\ref{pca_spots3} shows the marginal probability distribution for 
all relevant $\delta\Omega$ to have produced the data as derived from 
pairs with at least three common spots. The expectation value and the 
approximate $1\sigma$ confidence interval are $\delta\Omega=-0.048 \pm 
0.025$ d$^{-1}$. If we restrict the data to pairs with at least four 
common spots, we obtain the distribution shown in Fig.~\ref{pca_spots4}. 
The expectation value is $\delta\Omega=-0.045 \pm 0.026$ d$^{-1}$. The 
width of the distribution is slightly broader because fewer pairs define 
the result. However, since these combinations fix the unknowns better 
because of the larger number of spots, the increase in width is marginal.

The selection of pairs with five or more common spots delivers a very
wide marginal distribution with an uncertainty of more than 0.05 in
$\delta\Omega$. The number of pairs used is then 86.

The results are compatible with the differential rotation of the present 
Sun of $\delta\Omega= -0.0501$ d$^{-1}$ for a $\sin^2$-profile, according 
to Balthasar et al.\ (1986). The deviations of the expectation values 
based on Staudacher's observations from that value are not significant. 
Both probability distributions in Figs.~\ref{pca_spots3} and~\ref{pca_spots4} 
show a slight tendency to a smaller differential rotation in 1749--1799 
than today. This may be worth keeping in mind if more historical data 
are recovered, which would improve the statistics of the present work. 
Even if our study does not find a significantly different solar rotation 
profile, it is remarkable that these historical sunspot observations made 
by an amateur astronomer with a rather simple telescope reproduce the 
solar differential rotation nicely. The result is an indication for the 
quality of the drawings, which also led to the conclusion of an unusal 
butterfly diagram by Arlt (2009).

It may be worthwhile to check whether the average from the posterior 
distribution is a good measure of location for the differential rotation 
parameter. We find from simulated distributions of known differential
rotation that the average is indeed very close to the true value if the
posterior distribution becomes very small near the boundaries of the
considered parameter interval. This is not necessary (and is in fact not
true) for the individual distributions we obtained from individual pairs of 
drawings. The average of their average differential rotation parameters
is not equal to the average obtained from the total posterior, and thus
not close to the true value. All total posteriors shown in 
Figs.~\ref{pca_spots3}--\ref{pca_spots3_1762-1799} are extremely small close to
the interval boundaries of $-0.5$ and $0.5$.

An interesting experiment is the splitting of the set of observations into 
two smaller sets. Figure~\ref{pca_spots3_1749-1761} shows the marginal 
distribution for the drawings of 1749--1761, while Fig.~\ref{pca_spots3_1762-1799} 
shows the distribution for 1762--1799. The differential rotation parameters 
are $\delta\Omega=-0.073^{+0.047}_{-0.052}~{\rm d}^{-1}$ and 
$\delta\Omega=-0.039^{+0.027}_{-0.030}~{\rm d}^{-1}$, respectively, and 
indicate a decrease of the absolute value of the differential rotation 
during the second half of the 18th century. The difference is hardly 
significant though, because it is a $1\sigma$-effect.

A large differential rotation bears the possibility of a hydrodynamic 
shear instability, as studied first by Watson (1981), who found the 
solar shear to be nearly marginal. A 3D study with viscosity by 
Arlt et al. (2005) led to higher limits for the differential rotation, 
while higher powers of $\sin b$ may alter these results (Dziembowski 
\& Kosovichev 1987), which were not investigated here. If more data 
will become available and lead to a more accurate $\delta\Omega$, a 
test on its hydrodynamic stability would be required.

A change of differential rotation could also be important because 
the butterfly diagram has a different appearance during cycles~0 and 1 
(Arlt 2009). The equator was highly populated by spots and there is no 
clear migration direction of spot emergence. We can now use the spot 
latitudes determined as a side-product of the differential rotation 
determination and plot a butterfly diagram. Fig.~\ref{b_pca_20m_jpg} 
shows the positions of the 1512~spots used in this analysis. Since we 
can only plot the positions of spots associated with each other in 
pairs of drawings, their number is much smaller than in Arlt (2009). 
The main features of that earlier paper are also visible in the present 
graph: relatively normal cycles~3 and~4, while cycles~0 and~1 show a 
highly populated equator, which persists also into cycle~2. The 
appearance of the first two cycles of Staudacher's period somewhat 
resembles a topology that dynamo theorists call symmetric solutions 
of the magnetic field, in contrast to the anti-symmetric situation 
that we observe today. Symmetry here is meant with respect to the equator. 
Of course we do not have magnetic field polarities at hand for historical 
observations, and consequently the true topology of the dynamo mode will 
hardly ever be accessible.

Mean-field models typically show that symmetric and anti-symmetric 
solutions have very similar excitation thresholds (e.g.\ Chatterjee 
et al. 2004; Dikpati et al. 2004; Bonanno et al.\ 2005) and it is 
not straightforward to assume  the one should dominate the other 
all the time. A small change in $\delta\Omega$ might have favoured 
the symmetric solution in the 18th century, which exhibited a different 
butterfly diagram. Mean-field dynamo models with a back-reaction on the 
differential rotation tend to show excursions to symmetric geometry for 
{\em reduced\/} differential rotation though (e.g.\ Tobias 1997; 
K\"uker et al. 1999; Bushby 2006). Before we speculate too much, 
we conclude here that there is an indication for a change of 
$\delta\Omega$ during the 18th century, and we aim to stimulate 
future studies on the possible competition of the individual dynamo 
modes.

\acknowledgements
RA is grateful to the Deutsche Forschungsgemeinschaft for supporting
this work within the grant Ar355/6-1 and for the hospitality of the Kiepenheuer 
Institute for Solar Physics where the work has been initiated during 
a one-month visit.


\label{lastpage}

\end{document}